# Assessing foundational atomistic models for iron alloys under Earth's core conditions

Tianqi Wan[1#], Liangrui Wei[1#], Zepeng Wu[1], Renata M. Wentzcovitch[2-4], Yang Sun[1*]

[1]*Department of Physics, Xiamen University, Xiamen, 361005, China*
[2]*Department of Applied Physics and Applied Mathematics, Columbia University, New York, NY, 10027, USA*
[3]*Department of Earth and Environmental Sciences, Columbia University, New York, NY, 10027, USA*
[4]*Lamont–Doherty Earth Observatory, Columbia University, Palisades, NY, 10964, USA*

We assess the capability of recently developed foundational atomistic models (FAMs) to simulate iron alloys under the extreme pressures and temperatures of Earth's core. Static equations of state for hexagonal close-packed (hcp) and body-centered cubic (bcc) iron, computed using 17 FAMs, are benchmarked against *ab initio* calculations. Two representative models, MatterSim and MACE, are further evaluated for their ability to reproduce phonon spectra, liquid structure, and melting relations of iron at core conditions. While both models capture several key properties, MACE substantially overestimates the stability of bcc iron and fails to correctly describe the stability of hcp iron. Their performance is also examined for binary liquids, superionic phases, and a seven-component Fe–Ni–Si–S–O–H–C liquid. Although these FAMs were not explicitly trained on data from core conditions, they can reproduce several structural and dynamical properties across a wide range of compositions. However, none of the tested models consistently reproduces all first-principles benchmarks. By analyzing the origins of these discrepancies, we identify several limitations of current FAMs, particularly the lack of an explicit treatment of thermal electronic excitations, which significantly affect phase stability and thermodynamic properties under core conditions. We further discuss directions for improving FAMs to enable predictive simulations of core-forming materials under extreme conditions.

## I. Introduction

The Earth's core represents one of the most extreme conditions on the planet, with pressures up to 365 GPa and temperatures exceeding 5000 K [1]. The core consists primarily of iron, alloyed with small amounts of nickel and light elements [2-4]. Due to the extreme conditions, it is necessary to combine limited laboratory experiments with theoretical modeling of core-forming materials under seismic constraints to understand the core's structure and composition [5-15]. Thus, accurate atomistic simulations under extreme pressure–temperature conditions are essential for elucidating core processes ranging from heat transport at the core–mantle boundary [16] to inner-core crystallization [17,18].

For decades, *ab initio* molecular dynamics (AIMD) based on density functional theory (DFT) has been the primary approach for simulating materials under core conditions. Extensive studies have been conducted on pure Fe, including its equation of state [19-21], melting curve [8,22-26], thermal conductivity [16], viscosity [27], and seismic properties [28,29]. These simulations have also been extended to binary and ternary systems involving combinations of Fe with Ni, Si, S, O, C, and H [30-37]. However, the steep computational cost of DFT, which typically scales as $O(N^3)$ with system size N, severely restricts simulations to a few hundred atoms and picosecond timescales, thereby limiting its application to multicomponent core-forming materials.

Developments in machine-learning potentials (MLPs) have enabled large-scale simulations with near first-principles accuracy [38-42]. These MLPs are generally trained on DFT-generated datasets, allowing them to reproduce first-principles potential energy surfaces at substantially reduced computational cost. MLPs have been shown to provide satisfactory modeling of pure iron properties, including melting behavior and phase competition [43,44]. A few MLPs have also been developed to study thermodynamics [45-48], viscosity [49], and elasticity [50] in iron alloys. However, the applications of MLPs have so far been limited to binary or ternary systems due to the largely insufficient availability of DFT datasets for multicomponent core-forming iron alloys.

Recently, a new class of machine-learning models, referred to as foundational atomistic models (FAMs) or universal force fields, has emerged in materials science [51-62]. Unlike task-specific MLPs, FAMs are trained on extremely large and chemically diverse datasets spanning wide regions of compositional space across the periodic table, enabling transferable simulations across many material systems. Representative FAMs are trained on large-scale materials databases, including the Materials Project dataset [63] and the Open Materials 2024 database [64]. Recent benchmarking studies indicate that some FAMs can be applied to multicomponent material systems across broad chemical spaces [65-69]. However, most of these FAMs are developed using datasets generated under ambient or low-pressure conditions. Their performance in describing core-forming materials—particularly under extremely high-pressure and high-temperature conditions—remains unclear.

[#]These authors contribute equally.
[*]Email: yangsun@xmu.edu.cn

In this study, we systematically evaluate the performance of several state-of-the-art FAMs for modeling iron and iron alloys under Earth's core conditions. We first compute the static equations of state (EOS) for hexagonal close-packed (hcp) and body-centered cubic (bcc) iron using 17 different FAMs [51-58], benchmarking their ability to describe the compressional behavior of the most relevant crystalline phases in Earth's inner core. We then evaluate the FAMs based on iron's dynamical properties, including phonon dispersions, melting temperatures, and liquid structures, against ab initio data. These quantities provide key benchmarks for evaluating FAMs' ability to describe phase competition from both energetic and dynamical perspectives. The computed melting temperatures provide a direct measure of the high-temperature thermodynamics of the solid-liquid phase transition. We further extend the benchmark to chemically complex systems, including 6 Fe-X binary systems and a seven-component Fe–Ni–Si–S–O–H–C alloy relevant to Earth's core composition. We compare the pair correlation functions (PCFs) and mean square displacements (MSDs) for liquid and superionic phases between the FAMs and ab initio data. The PCFs provide rich information on the local coordination environment and atomic mean forces, whereas the MSDs characterize the intrinsic atomic diffusion properties. Together, these benchmarks constitute a comprehensive assessment of general-purpose FAMs for Earth-core materials under extreme pressure and temperature conditions, evaluating their ability to capture the key structural, kinetic, and thermodynamic properties governing phase behavior and providing guidance for future development.

## II. Methods

### A. Foundational atomistic models

All simulations involving FAMs were performed using the Atomic Simulation Environment (ASE) [70], with each FAM implemented as a calculator to evaluate energy, atomic force, and pressure. For pure iron, static optimizations were carried out to determine the equilibrium $c/a$ ratio of hcp Fe at various volumes using the Broyden–Fletcher–Goldfarb–Shanno (BFGS) optimizer within ASE. Phonon dispersions for bcc and hcp Fe were computed using the finite-displacement approach with PHONOPY [71,72], where atomic forces were provided directly by the corresponding FAMs. Molecular dynamics (MD) simulations were conducted in the canonical (NVT) ensemble using a Nosé–Hoover thermostat implemented in ASE [73-75]. For iron alloys, supercells containing 250–360 atoms were employed to model the liquid, superionic bcc, and superionic hcp phases, the same as those used in the AIMD calculations. FAM-MD simulations were carried out with a time step of 0.5 *fs*. For different binary alloys and the seven-component Fe–Ni–Si–S–O–H–C liquid, MD simulations were performed for up to 100 *ps* to ensure converged physical properties for each FAM.

### B. Melting temperature calculation

The melting temperature of iron is based on the condition where the free energy difference between solid and liquid is zero, i.e., $\Delta G_{\mathrm{FAM}}^{L-S}(T_m^{FAM}) = 0$ . Here, the free energy difference between solid and liquid iron for a FAM was calculated using Hamiltonian thermodynamic integration (TI) within the framework developed in Ref. [25]. The reference system is described by an embedded-atom-method (EAM) potential. Then the liquid-solid Gibbs free-energy difference for the FAM can be obtained [25] via

$$\Delta G_{\mathrm{FAM}}^{L-S}(T) = f_{PV}(T) + f_{TI}(T) + \Delta G_{\mathrm{EAM}}^{L-S}(T) \qquad (1)$$

where $f_{PV}(T) = P(V_{FAM}^L - V_{FAM}^S) - P(V_{\mathrm{EAM}}^L - V_{\mathrm{EAM}}^S) - \left(\int_{V_C^L}^{V_{FAM}^L} P_{\mathrm{EAM}}^L(V)dV - \int_{V_{\mathrm{EAM}}^S}^{V_{FAM}^S} P_{\mathrm{EAM}}^S(V)dV\right)$. Here $V_{\mathrm{FAM}}^L$ (or $V_{\mathrm{FAM}}^S$) and $V_{\mathrm{EAM}}^L$ (or $V_{\mathrm{EAM}}^S$) are the equilibrium volumes of the liquid (or solid) at given $(P, T)$ for the FAM and EAM, respectively. The equilibrium volumes of the FAM were obtained from NPT-MD simulations. $P_{\mathrm{EAM}}^L(V)$ and $P_{\mathrm{EAM}}^S(V)$ are the equations of state of liquid and solid for EAM, respectively. The TI term, written as $f_{TI} = F_{TI}^L(V_{\mathrm{FAM}}^L, T) - F_{TI}^S(V_{\mathrm{FAM}}^S, T)$ accounts for the TI difference between the liquid and solid states. Here, $F_{TI}^L(V_{\mathrm{FAM}}^L, T)$ and $F_{TI}^S(V_{\mathrm{FAM}}^S, T)$ are the Helmholtz free energy difference between the FAM and EAM for liquid and solid, respectively, which are computed by TI as implemented in ASE [70]. In TI simulation, the force acting on each atom was coupled by $f = (1-\lambda)f_{\mathrm{EAM}} + \lambda f_{\mathrm{FAM}}$. A Nosé–Hoover thermostat was used to maintain the target temperature, and a time step of 1.0 fs was used to integrate Newton's equations of motion in the TI-MD. The reference free energy term, $\Delta G_{\mathrm{EAM}}^{L-S}(T)$ , was computed in Ref. [25].

### C. First-principles calculations

DFT calculations were performed using the Vienna *Ab initio* Simulation Package (VASP) code [76,77], employing the projector-augmented-wave (PAW) method [78] in conjunction with the Perdew–Burke–Ernzerhof (PBE) exchange–correlation functional [79]. The electronic entropy was included using the Mermin functional [80,81], with the electronic temperature set equal to the ionic temperature. Supercells containing 250–360 atoms were used to model liquid, superionic bcc, and superionic hcp phases. A time step of 1 *fs* was used for the Fe–O, Fe–C, Fe-Si, Fe-S, Fe-Ni binary systems and Fe-Ni-Si-S-O-H-C seven-element systems, while a smaller time step of 0.5 fs was employed for the Fe–H system to ensure numerical stability. PAW potentials with valence configurations of $3d^8 4s^2$ (Ni), $3d^7 4s^1$ (Fe), $3s^2 3p^4$ (S), $3s^2 3p^2$ (Si), $2s^2 2p^4$ (O), $2s^2 2p^2$ (C), and $1s^1$ (H) were used. A plane-wave energy cutoff of 400 eV and $\Gamma$-point sampling were applied in the AIMD simulations. These settings are sufficiently accurate for calculations of iron's liquid structure and equation of state [24]. For the comparison of iron's melting temperature, we used previously computed results obtained with the $3s^2 3p^6 3d^7 4s^1$ (PAW16) potential for iron, as reported in Ref. [25]. Phonon calculations were performed using the finite-displacement method as implemented in the PHONOPY package [71,72], with 144-atom supercells for hcp Fe and 64-atom supercells for bcc Fe.

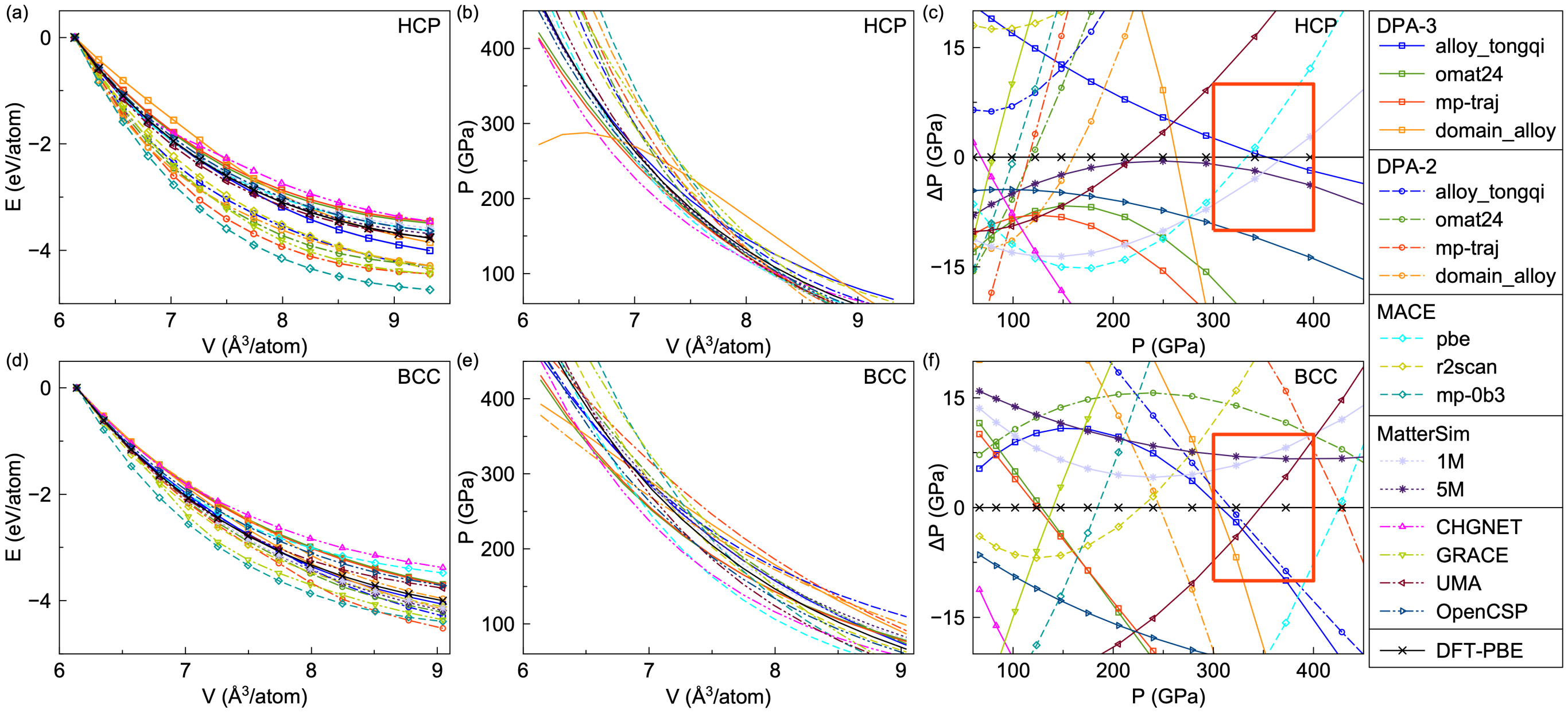

**FIG. 1** Static equations of state for (a-c) hcp and (d-f) bcc iron computed by DPA-3 [53], DPA-2 [52], MACE [54], MatterSim [58], CHGNet [51], OpenCSP [55], UMA [56], GRACE [57], and DFT. Energy $E$ is referenced to the data of the smallest volume in (a) and (d).

## III. Results

### A. Pure iron

Figure 1 presents the static equations of state (EOS) for iron computed with 17 different FAMs [51-58] and benchmarked against DFT calculations with the PBE functional. The energy–volume ($E$–$V$) curves for hcp shown in Fig. 1a exhibit substantial variations among different FAMs, indicating a strong dependence on the architecture of FAMs. Even within the same FAM family, models trained on different datasets can yield markedly different predictions. For instance, among the DPA-3 models [53], those trained on alloy datasets (DPA-3_alloy_tongqi [82] and DPA-3_domain_alloy [83]) show better agreement with DFT compared to models trained on the OMat24 dataset [64] (DPA-3_omat24) or the Materials Project trajectory dataset [63] (DPA-3_mp-traj). Among the MACE models [54], MACE-PBE generally outperforms MACE-mp-0b3 and MACE-r2SCAN. The discrepancy with MACE-r2SCAN is expected, as it is trained using a different exchange–correlation functional. However, even though MACE-mp-0b3 is also trained using the PBE functional, its poorer performance compared to MACE-PBE is likely due to differences in the training data. In particular, MACE-mp-0b3 is trained on the Materials Project trajectory database [63], whereas MACE-PBE uses the MatPES dataset [84]. The two MatterSim versions [58] yield similar results, with MatterSim-5M showing slightly better agreement with DFT.

Comparisons between Figs. 1a and 1d reveal that the FAMs exhibit similar performance for both hcp and bcc phases. By fitting the $E$–$V$ curves to the third-order Birch–Murnaghan EOS, we derived the pressure–volume ($P$–$V$) relationships depicted in Figs. 1b and 1e. Analysis of the compression curves and pressure deviations with DFT calculations ($\Delta P$ vs. $P$) for hcp-Fe shows that several models, including DPA-3-mp-traj, DPA-2-mp-traj, CHGNet [51], MACE-mp-0b3, OpenCSP [55], and UMA [56], perform accurately within the P<150 GPa range ($\Delta P < 15$ GPa). However, their precision deteriorates markedly under Earth's inner core conditions ($P > 330$ GPa), where UMA [56] reaches an error of ~40 GPa, while DPA-2-mp-traj and MACE-mp-0b3 exhibit substantial deviations approaching 200 GPa. In contrast, only MatterSim-1M, MatterSim-5M, and MACE-PBE maintain robust performance ($\Delta P < 15$ GPa) across the entire pressure regime, while DPA-3_alloy_tongqi excels specifically at higher pressures.

A similar trend is observed for bcc-Fe in Fig. 1e and 1f. FAMs, including GRACE [57], MACE-r2SCAN, DPA-3-mp-traj, and DPA-3-omat24 variants demonstrate high fidelity up to 150 GPa with $\Delta P < 10$ GPa. However, they exhibit a drastic loss of accuracy, with deviations reaching 80 GPa, when extended to the core-relevant regime ($P > 330$ GPa). In contrast, MatterSim-1M, MatterSim-5M, and DPA-2-omat24 maintain good performance with $\Delta P < 15$ GPa throughout the entire pressure range. MACE-PBE, while showing reliable alignment at pressures exceeding 300 GPa, performs less optimally in the lower-pressure regime. Conversely, CHGNet fails to accurately capture the $P$–$V$ response of bcc-Fe across the full range of pressures considered. To quantitatively assess model reliability, we defined a performance threshold indicated by the red rectangles in Figs. 1c and 1f. A FAM is considered applicable for inner core conditions (300-400 GPa) if the pressure deviation remains within $\pm 10$ GPa of the DFT data. Consequently, MACE-PBE, MatterSim-1M, MatterSim-5M, and DPA-3_alloy_tongqi emerge as the most promising candidates, as they uniquely satisfy our reliability criteria for both hcp and bcc phases. Due to the absence of training data for key light elements of H, C, and O in the DPA-3_alloy_tongqi model, we restrict our subsequent analysis to MACE-PBE (hereafter referred to as MACE), MatterSim-1M (referred to as MS-1M), and MatterSim-5M (referred to as MS-5M).

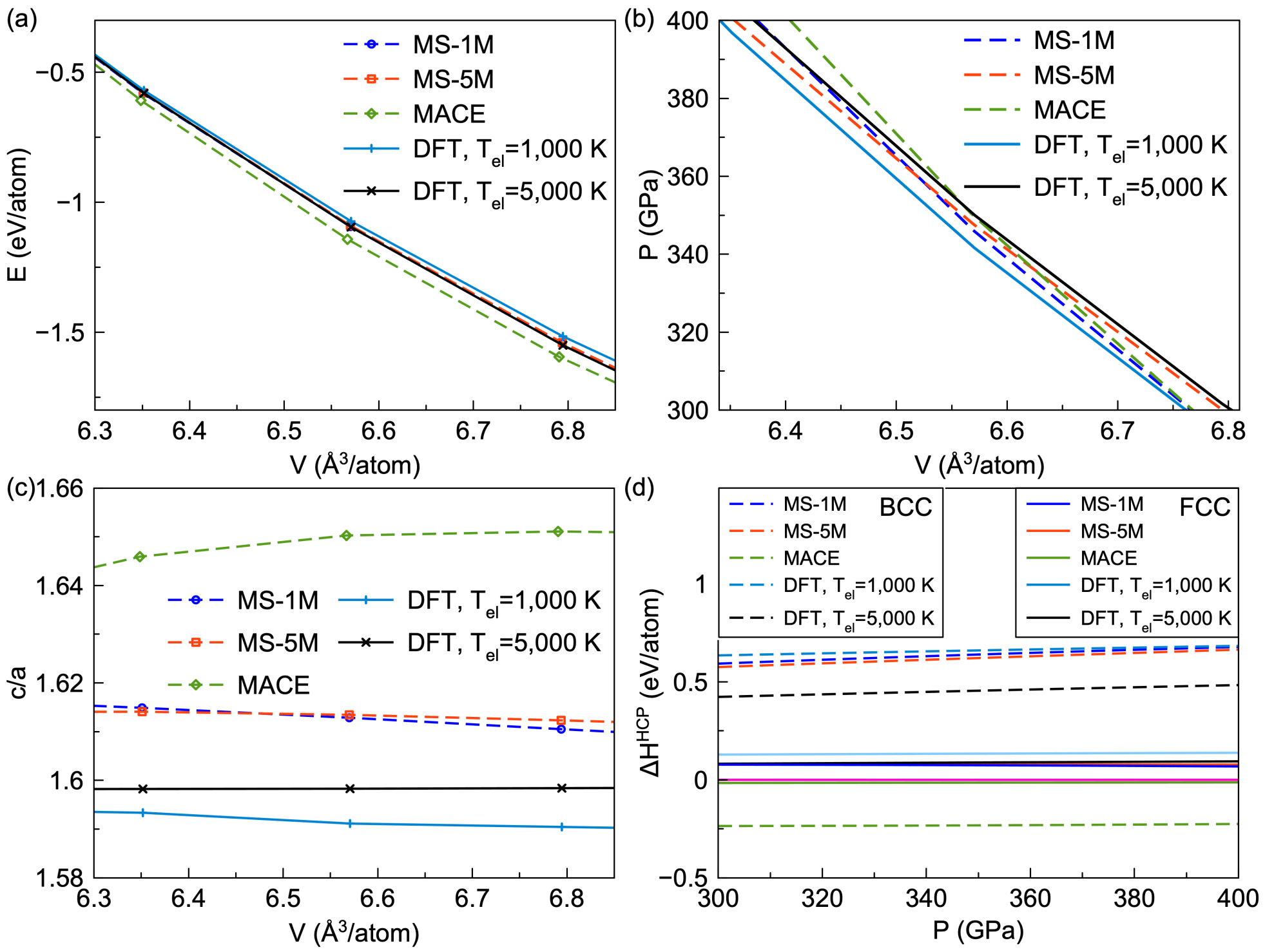

**FIG. 2** Comparison among MS-1M, MS-5M, MACE and DFT for pure Fe phases under Earth's core conditions. (a) E–V curves of hcp-Fe. (b) Pressure–volume relations of hcp-Fe. (c) The c/a ratio of hcp-Fe. (d) Relative enthalpy with respect to hcp-Fe.

Next, we present a detailed benchmark of the three selected FAMs against DFT calculations across a pressure range of 300–400 GPa, corresponding to the volume range of ~6.3-6.8 $Å^3/\mathrm{atom}$ based on the DFT EOS. The $E$–$V$ curves (Fig. 2a) demonstrate that both MS-5M and MS-1M capture the DFT potential energy surface with high fidelity, maintaining a deviation below 10 meV/atom. In contrast, MACE shows a larger systematic deviation of about 50 meV/atom. While these FAMs do not explicitly incorporate electronic entropy contributions, the impact of electronic temperature in DFT, which can shift E–V curves by ~30 meV/atom between $T_{el}$=1000 K and $T_{el}$=5000 K, is comparable to the observed deviations between MS models and DFT. The compression curves are compared in Fig. 2b. Both MS-1M and MS-5M yield a decent match with the DFT EOS across the entire pressure regime, with deviation of ~ 5 GPa. MACE exhibits a pronounced slope mismatch relative to the DFT benchmark within the 300–450 GPa range, indicating a failure to capture the correct compressibility. The $c/a$ ratio of hcp Fe computed with different FAMs are compared in Fig. 2c. DFT predicts $c/a$ values in the range of 1.59–1.60, depending on the electronic temperature. The MS-1M and MS-5M yield a slightly higher value of 1.61–1.62. MACE shows a larger deviation with $c/a$ exceeding 1.64.

Figure 2(d) establishes the benchmark for phase competition among solid Fe structures at 300–400 GPa. Using the enthalpy of hcp Fe as the reference, both MS-1M and MS-5M correctly predict hcp-Fe as the most stable phase, followed by fcc-Fe and then bcc-Fe ($H_{\mathrm{bcc}} > H_{\mathrm{fcc}} > H_{\mathrm{hcp}}$), which is consistent with the trend of the DFT calculation. For bcc-Fe, the results of MS-1M and MS-5M are closer to DFT data computed with an electronic temperature of 1000 K. For fcc-Fe, they are closer to the one computed with an electronic temperature of 5000 K. MACE completely fails to reproduce the stability sequence obtained in the DFT data. It identifies fcc as having an enthalpy very close to that of the hcp phase, and bcc-Fe as the lowest-enthalpy phase. Thus, the static enthalpy results show that MS-1M and MS-5M capture the relative phase stability predicted by DFT, whereas MACE substantially overestimates the enthalpic stability of bcc Fe. This pronounced overestimation of bcc stability highlights MACE's limitation in resolving phase competition in iron under inner-core conditions.

Having established the phase ordering based on the static enthalpy, we next examine whether these FAMs also reproduce the relative dynamical stability of bcc and hcp Fe at 323 GPa. As shown in Fig. 3a and 3b, the phonon spectra bcc-Fe described by MS-1M and MS-5M exhibit strong imaginary phonon modes, similar to the DFT calculation. This instability is a well-known feature of bcc Fe at zero Kelvin and inner core pressures [85]. In contrast, MACE predicts a dynamically stable bcc phonon spectrum, with no imaginary frequencies throughout the Brillouin zone in Fig. 3c. The situation is reversed for hcp-Fe. DFT calculations indicate that the hcp phase remains dynamically stable across the entire Brillouin zone. The phonon spectra from MS-1M and MS-5M calculations reproduce this behavior and show good agreement with DFT results. However, the phonon spectrum computed by MACE shows large differences with DFT results and even unstable imaginary phonon modes in hcp-Fe. The phonon calculations, therefore, provide independent dynamical evidence consistent with the enthalpy results in Fig. 2(d). Taken together, Fig. 2(d) and Fig. 3 show that MS-1M and MS-5M reproduce both the static energetic ordering and the relative dynamical stability of bcc and hcp Fe, whereas MACE fails in both respects by overestimating the stability of bcc Fe relative to hcp Fe.

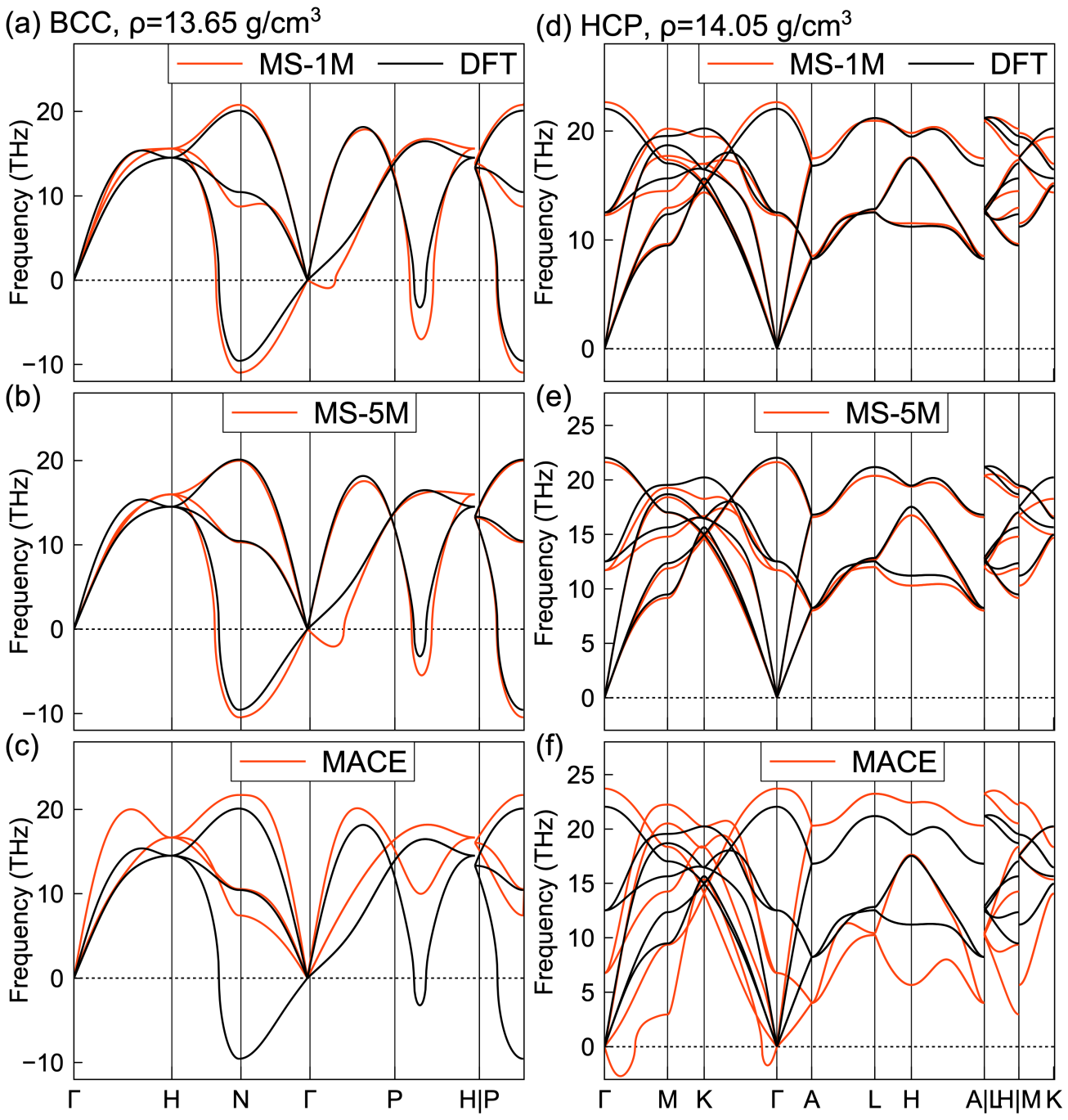

**FIG. 3** Phonon dispersion for (a-c) bcc and (d-f) hcp iron at 323 GPa from FAM and DFT calculations.

Despite pronounced differences in solid-phase description, all three models—MS-1M, MS-5M, and MACE—successfully reproduce the PCF of liquid iron at 6000 K and a density of 13.135 $g/cm^3$, corresponding to 323 GPa based on AIMD, as shown in Fig. 4. While all three FAMs accurately predict the position of the first peak, they all slightly overestimate the height of the first peak, with MACE yielding results closest to the DFT data. Thus, the liquid structure of pure Fe is reasonably well captured by these models.

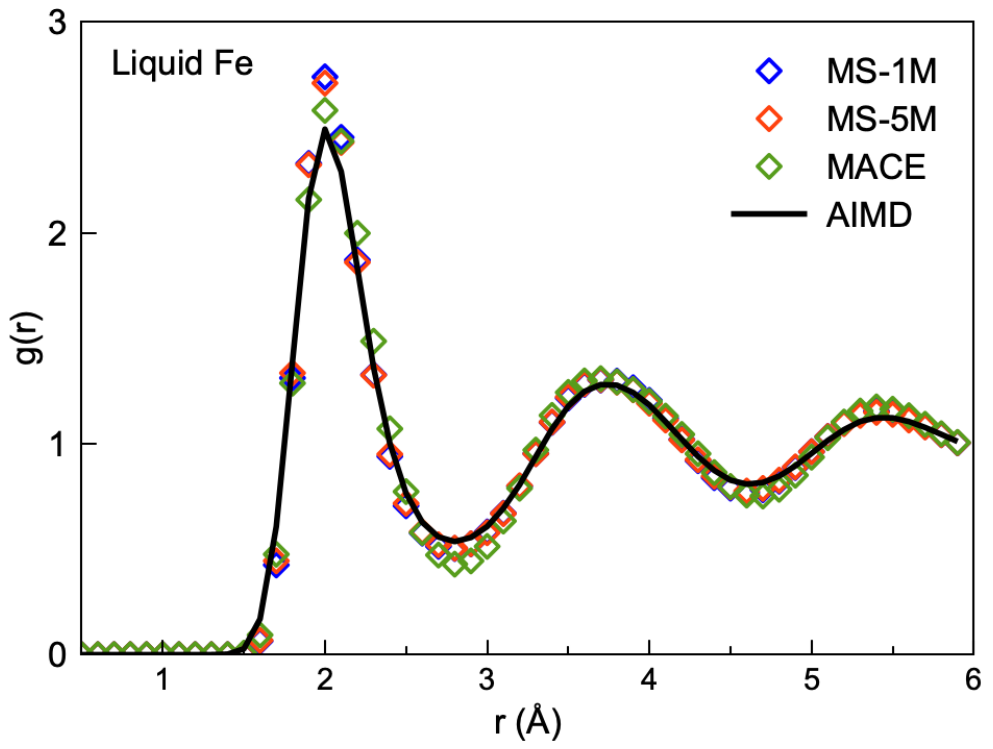

**FIG. 4** PCF of pure liquid iron at 6000 K and a density of 13.135 $g/cm^3$.

We further compute the melting temperatures of the bcc, hcp, and fcc phases of Fe for MS-5M and MACE at 323 GPa and 360 GPa. The resulting melting temperatures are compared with previously reported AIMD results [25], as summarized in Table 1. MS-5M reproduces the expected relative stability among the three phases, but it systematically over-stabilizes hcp and fcc w.r.t. the liquid compared with AIMD data, resulting in higher melting temperatures. In contrast, MACE fails to reproduce the correct phase ordering and strongly over-stabilizes bcc-Fe. Moreover, we are unable to determine the melting temperature of hcp iron for MACE because the hcp structure spontaneously transforms into bcc during the TI calculation. This behavior is consistent with static calculations, in which MACE predicts the bcc phase to be much more stable than hcp in Fig. 2d.

**Table 1** Melting temperature of hcp, fcc, and bcc phases based on the MS-5M and MACE, compared to AIMD data from Ref. [25, 36]. "—" in the MACE indicates the instability of the hcp phase in the free-energy calculations. "NA" in the AIMD results indicates a lack of available data.

| Method | P/GPa | $T_m^{hcp}$/K | $T_m^{fcc}$/K | $T_m^{bcc}$/K |
|---|---|---|---|---|
| MS-5M | 323 | 7045 | 6777 | 6100 |
| | 360 | 7416 | 7142 | 6417 |
| MACE | 323 | — | 6582 | 7313 |
| | 360 | — | 6840 | 7678 |
| AIMD (PAW16) | 323 | 6357 | 6226 | 6168 |
| | 360 | 6692 | NA | 6519 |

## B. Iron alloys

In addition to Fe, Earth's core contains small amounts of Ni and light elements, including H, C, O, Si, S [1]. The broad elemental coverage is the key advantage of FAMs. Therefore, we assess the capability of FAMs to describe iron alloy systems under the extreme conditions of the Earth's core. Since MS-1M, MS-5M, and MACE provide a reasonable description of pure Fe under core conditions, we focus on these three FAMs to test iron alloys. We focus on liquid partial PCFs because they directly reflect species-resolved local structure and effective interatomic interactions related to the mean force in statistical theory [86]. We also evaluate MSDs to test whether the models reproduce light-element diffusion, especially in superionic phases.

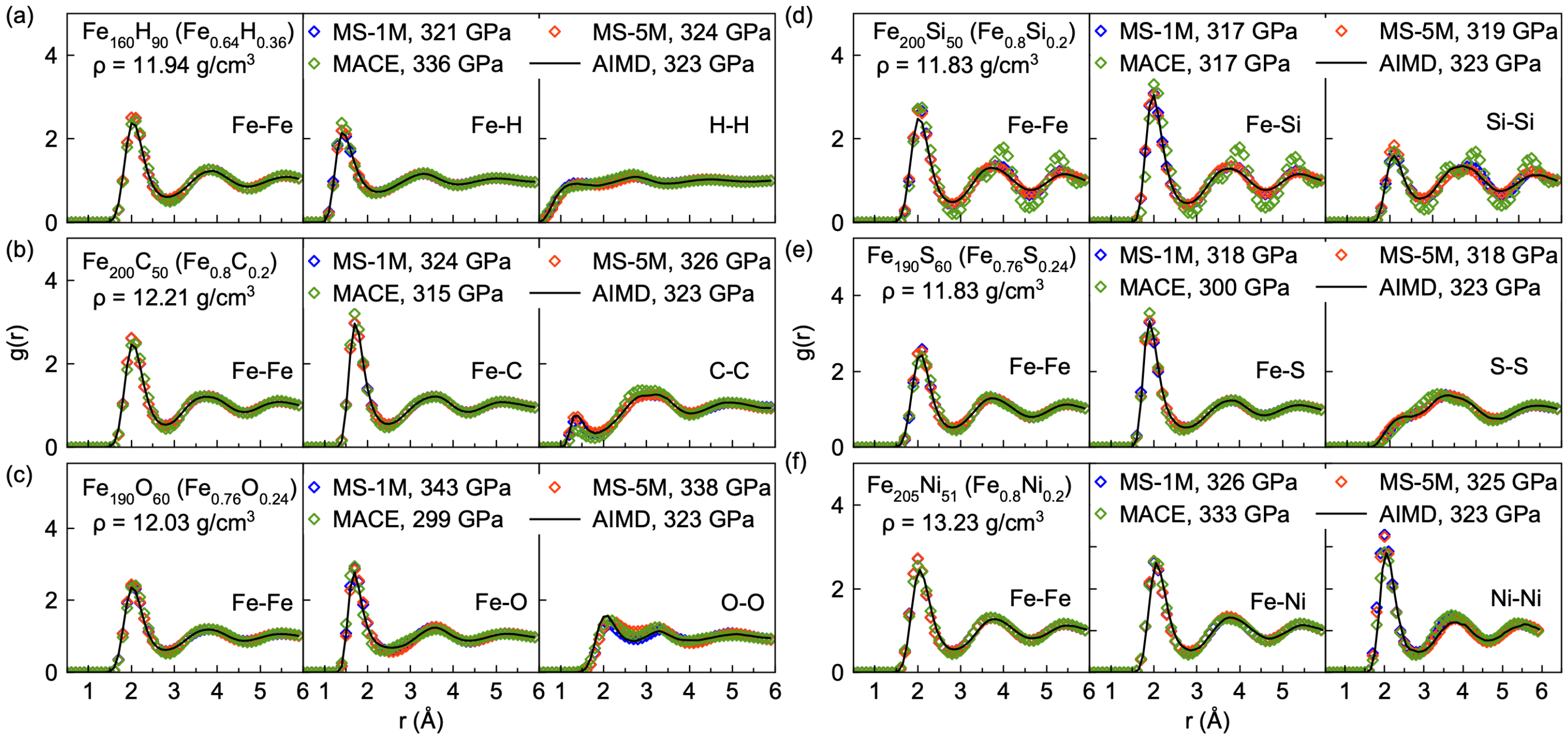

**FIG. 5** Comparison of PCFs for binary iron alloy melts at 5,800 K. (a) $Fe_{160}H_{90}$. (b) $Fe_{200}C_{50}$. (c) $Fe_{190}O_{60}$. (d) $Fe_{200}Si_{50}$. (e) $Fe_{190}S_{60}$. (f) $Fe_{205}Ni_{51}$ at 6,000 K. Simulations with different models are performed at the same density, corresponding to 323 GPa in the AIMD simulation. The resulting pressures from the different FAM simulations are listed in the legend.

First, we perform a systematic assessment of binary melts. Since the results are compared with AIMD, the simulation size of these binary liquids is limited to ~250 atoms. We set the Ni content or the light-element content to 20–36% to collect sufficient statistics on pair interactions. As shown in Fig. 5, the PCFs predicted by the tested FAMs mostly agree with the AIMD data, indicating that these FAMs can reasonably describe the coordination environment and characteristic interatomic distances. However, MACE lacks the ability to describe the Fe–Si system. In Fig. 5d, the Fe–Si PCFs calculated by MACE display systematically sharper peaks than those obtained from DFT calculations. They also show peak splitting in the second PCF for all three pairs, suggesting the presence of artificial medium-range order in the simulated Fe–Si melts. The performances of MS-1M and MS-5M are similar, and both provide satisfactory descriptions of these binary melts, comparable to DFT calculations.

While MS-1M and MS-5M generally provide satisfactory descriptions of the compositions tested in Fig. 5, they fail to accurately simulate the Fe–O system at reduced oxygen content. In Fig. 6, simulations of $Fe_{220}O_{30}$ (12 at.%) melts show a series of artificial results from simulations with MS-1M and MS-5M. The PCFs of both Fe–O and O–O pairs simulated by MS-1M deviate significantly from the AIMD simulations. It forms short O–O bonds of ~1.2 Å and exhibits much higher peaks than those obtained from AIMD. For MS-5M, the Fe–O PCF becomes closer to the AIMD data, but the O–O pair still exhibits a much higher peak.

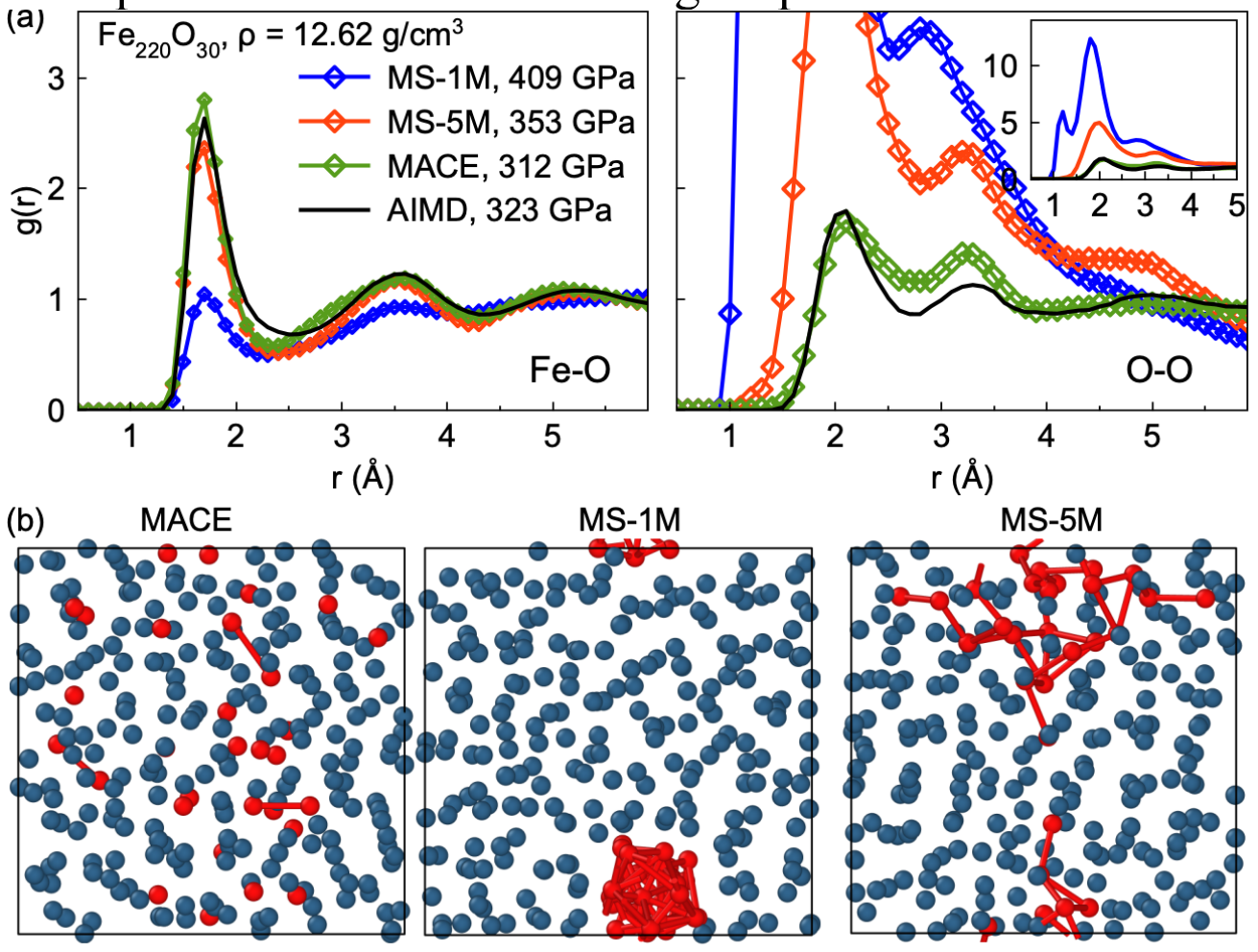

**FIG. 6** Liquid structure of $Fe_{220}O_{30}$ ($x_O$=12 at.%) at 5,800 K and a density of 12.62 g/cm$^3$, corresponding to 323 GPa in the AIMD. (a) Partial PCF of Fe–O and O–O. The pressures obtained from the different FAM simulations are listed in the legend. (b) Atomic configurations for MACE, MS-1M, and MS-5M after 15 ps of simulation, starting from a uniform O distribution. O-O bonds are visualized using a cutoff of 2.1 Å.

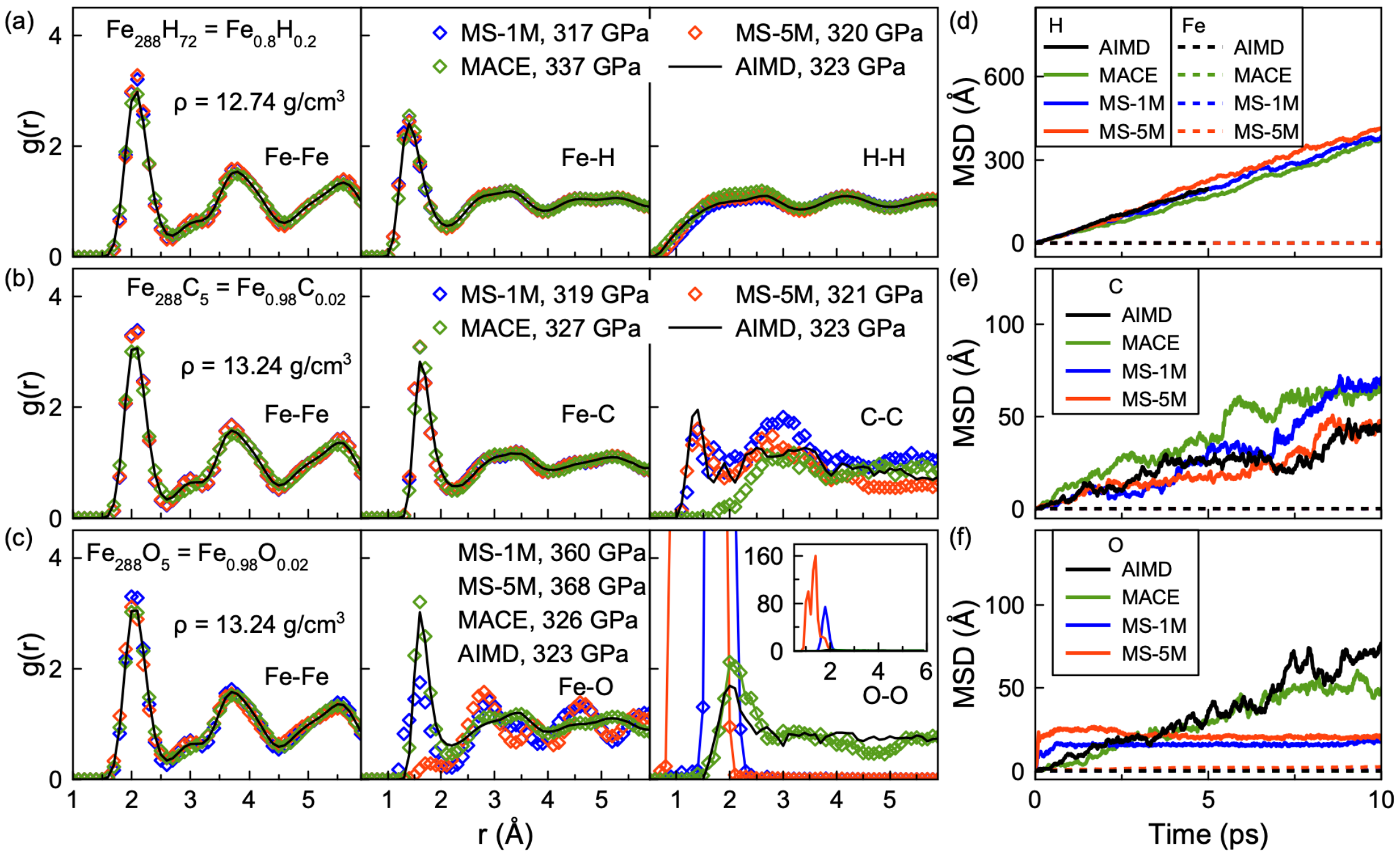

**FIG. 7** The PCF of the superionic hcp phase (a) $Fe_{288}H_{72}$, (b) $Fe_{288}C_5$, and (c) $Fe_{288}O_5$. (d-f) Mean square displacement (MSD). Simulations with different models are performed at the same density, corresponding to 323 GPa in the AIMD. The pressures obtained from the different FAM simulations are listed in the legend.

The atomic structures from the MD trajectories are shown in Fig. 6b. Starting from the same configuration with a uniform oxygen distribution, the behaviors diverge markedly after 15 ps of simulation. MACE preserves a uniform oxygen distribution consistent with the AIMD results, whereas MS-1M exhibits strong phase segregation, with oxygen atoms spontaneously clustering into dense aggregates. MS-5M shows a more uniform distribution than MS-1M, but still displays localized oxygen enrichment. Such unrealistic segregation and short O–O bond formation are concentration-dependent, as the system with a higher O concentration exhibits more regular behavior in Fig. 5. The comparisons among different models were performed at the same temperature and density as the corresponding AIMD simulations. The pressures predicted by the FAMs reflect their accuracy in describing the EOS for the corresponding alloy composition. For most binary liquids, the predicted pressures remain close to the AIMD values (Fig. 5), with deviations of <5 GPa for MS-1M/5M and ~10 GPa for MACE. In dilute $Fe_{220}O_{30}$, however, the oxygen segregation predicted by MS-1M/ 5M is accompanied by pressure deviations exceeding 30 GPa, indicating failures in both liquid structure and EOS.

H, C, and O are superionic in the hcp phase under Earth's inner-core conditions, a key feature revealed by AIMD simulations [33]. In the superionic state, iron atoms exhibit solid-like vibrational motion, while the light elements undergo liquid-like diffusion within the hcp lattice. To benchmark the accuracy of FAMs in describing the superionic state, we simulated the hcp phases of $Fe_{0.8}H_{0.2}$, $Fe_{0.98}C_{0.02}$, and $Fe_{0.98}O_{0.02}$ and compared the results with AIMD simulations. We chose small concentrations of C and O because of their small solid–liquid partition coefficients [37,87]. Figure 7a shows that MS-1M, MS-5M, and MACE all provide consistent descriptions of the Fe–H system compared with AIMD data in terms of structure. Moreover, as shown by the MSD in Fig. 7d, the kinetics of superionic Fe–H are well reproduced by all three FAMs relative to AIMD. For the Fe–C system (Fig. 7b), MACE fails to correctly describe the C–C PCF, completely missing the first peak. MS-1M overestimates the intensity of the second C–C peak, while MS-5M performs better. The MSDs from all three FAMs agree with AIMD results (Fig. 7e). For Fe–O, MS-1M and MS-5M show inconsistent structures in Fig. 7c. The O–O PCFs do not agree with AIMD, which is again associated with the formation of O aggregates, as observed in $Fe_{0.88}O_{0.12}$ melts. The MSDs of O from MS-1M and MS-5M show a plateau because the O atoms do not diffuse once they form aggregates. MACE provides a good description of superionic Fe–O using both PCF and MSD, as shown in Figs. 7e and 7f. The inability to fully reproduce the superionic behavior of light elements can be a significant limitation for these FAMs in simulating the inner core.

To evaluate the performance of the three FAMs in describing multicomponent systems, we further extended our benchmarks to a chemically complex seven-element iron melt, $Fe_{150}Ni_{15}Si_{25}S_{15}H_{30}O_{80}C_{15}$, a composition within current estimates of Earth's outer core composition [2]. Using the same volume as in AIMD, MS-5M reproduces the pressure well, giving 320 GPa, whereas MACE underestimates it as

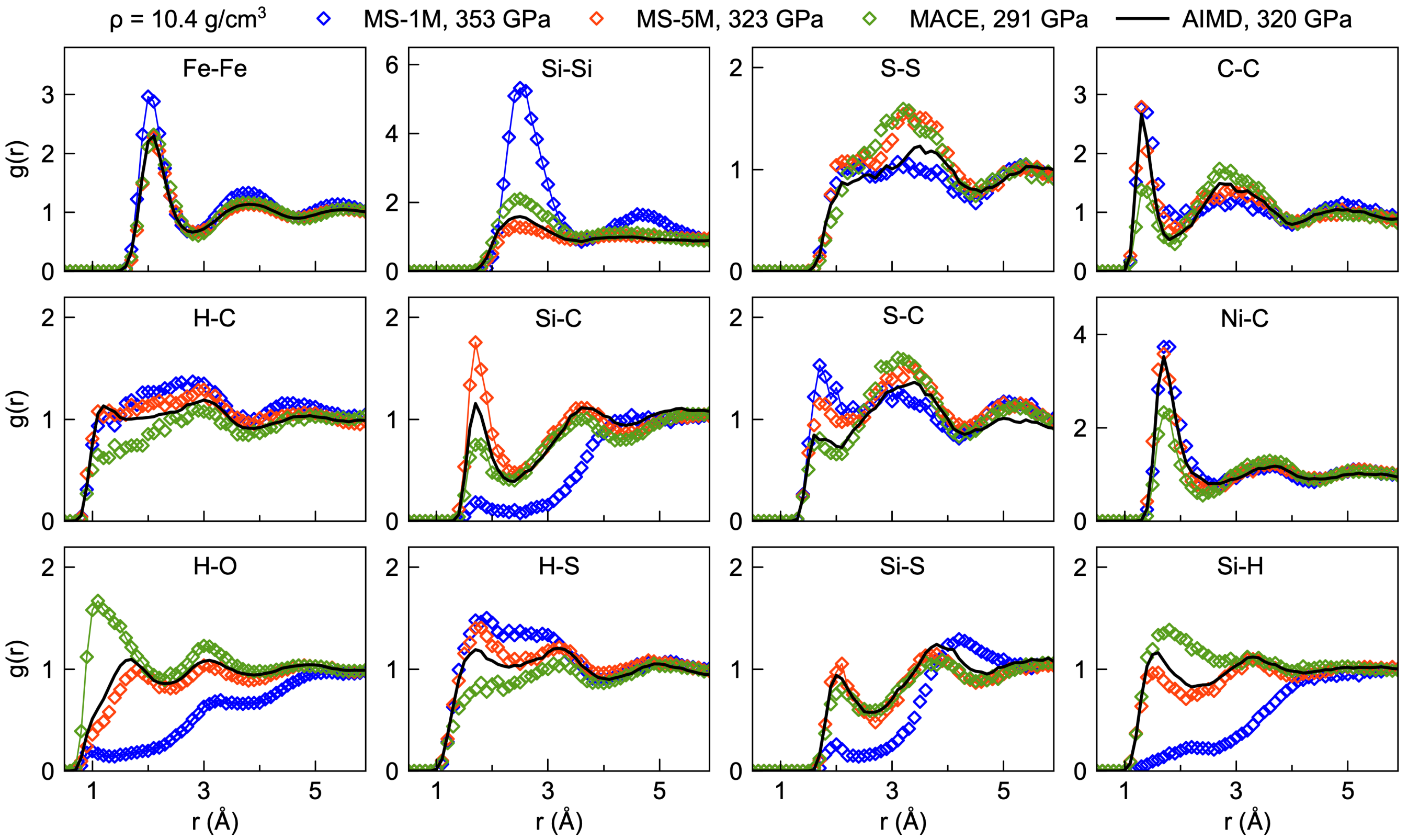

**FIG. 8** Representative partial PCFs for $Fe_{150}Ni_{15}Si_{25}S_{15}H_{30}O_{80}C_{15}$ melts at 5800 K, with a density corresponding to 320 GPa in the AIMD simulation. Simulations with different FAMs are performed at the same density, and the resulting pressures are listed in the legend.

291 GPa and MS-1M overestimates it as 353 GPa. Figure 8 presents 12 representative pairs out of the 28 distinct partial PCFs for this system, highlighting the main deviations with AIMD. The complete set of 28 partial PCFs and the corresponding MSDs are provided in Appendix Fig. A1. The comparison shows that model transferability is strongly pair-dependent. MS-5M gives the most balanced agreement with AIMD, with small deviations for Si-Si, S-S, Si-C, and H-S pairs. MACE exhibits broader discrepancies for Si-Si, S-S, C-C, H-C, Si-C, Ni-C, H-O, Si-S, and Si-H. MS-1M generally has much larger errors. Thus, for multicomponent core-forming melts, the tested FAMs do not yet reproduce pair correlations and interatomic interactions with sufficient ab initio accuracy. The MSDs of this multicomponent system were also evaluated over a 10-ps MD trajectory in Appendix Fig. A1. Three FAMs reproduce the diffusive behavior of Fe, Ni, and the light elements in good agreement with the AIMD data, indicating quantitatively accurate diffusion coefficients.

## IV. Discussion

We have systematically assessed the performance of representative FAMs for iron and iron-rich alloys under Earth's core conditions. For pure Fe, MS-1M and MS-5M reproduce the EOS and relative enthalpy difference between hcp and bcc phases up to inner-core pressures, whereas several other models degrade at high pressures. MACE also describes the EOS reasonably well but substantially overestimates the stability of bcc Fe, resulting in an inability to reproduce the correct static phase relations and phonon behavior. MS-1M, MS-5M, and MACE capture the liquid structure of Fe reasonably well; however, discrepancies emerge in melting relations. MS-5M preserves the correct phase ordering but over-stabilizes hcp relative to the liquid, resulting in an overestimated melting temperature by a few hundred Kelvin. MACE fails to describe the hcp stability. These inconsistencies likely reflect differences in Fe training datasets and DFT settings, which can strongly affect Fe's phase stability under core conditions [25]. For instance, MACE was trained using the PAW14 potential for iron [54], which has been shown to yield different temperatures than the PAW16 potential, which includes additional 3s electrons [26].

For iron alloys, model-dependent deficiencies become more pronounced. MS-1M and MS-5M generally reproduce binary liquid structures but show unphysical oxygen aggregation in dilute Fe–O systems and fail to recover the correct superionic behavior of O. MACE performs better for oxygen-bearing systems but exhibits structural artifacts in Fe–Si melts. In the seven-element melt, MS-5M and MACE maintain overall structural and dynamical consistency with AIMD, whereas MS-1M shows noticeable degradation in liquid structure simulations. These results regarding light-element behavior in Fe's hcp and liquid phases demonstrate that current FAMs are not yet universally transferable and that predictive fidelity for core-forming materials is highly sensitive to chemical composition and bonding environment.

Several factors may contribute to the observed discrepancies. First, most FAMs are primarily trained on ambient-pressure data, with limited representation of ultrahigh-pressure environments [58]. These extreme conditions can significantly modify bonding character, especially for oxygen in iron, where it transforms from ionic to metallic behavior under compression [88-91]. Second, training datasets may not adequately sample dilute light-element configurations relevant to core compositions in our test cases. Moreover, current FAMs do not explicitly incorporate electronic entropy or temperature-dependent electronic free-energy contributions, which can influence phase stability at inner-core conditions [44]. Nevertheless, these FAMs, especially the MS models and MACE, already provide a good starting point for developing machine-learning models that can cover these seven elements, or more, under Earth's core conditions with further fine-tuning and additional data.

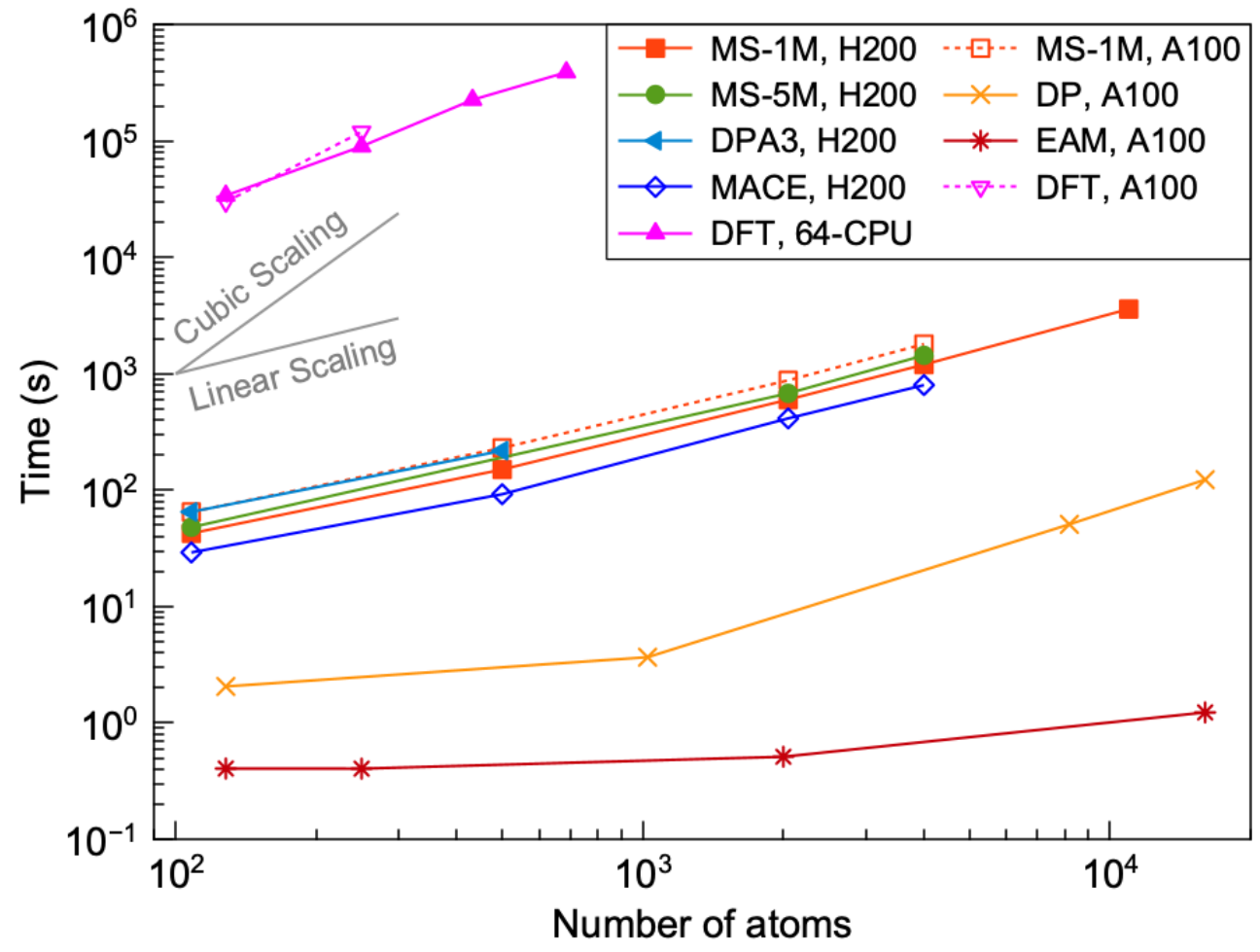

**FIG. 9** Comparison of wall time to perform a 1-ps MD simulation as a function of atom number for hcp-Fe at 6,000 K and 323 GPa, using NVIDIA H200, A100 GPU and a dual-socket Intel Xeon Gold 6338 CPU node (64 physical cores).

To evaluate the efficiency of FAMs for simulating core-forming materials, we benchmarked their computational performance against DFT, a specialized neural-network-based Deep Potential (DP) model [44], and a semi-empirical embedded-atom-method (EAM) potential [17] in Fig. 9. AIMD remains the most computationally demanding method and exhibits cubic scaling $O(N^3)$. FAMs such as MS-1M, MS-5M, MACE, and DPA-3 reduce computational cost by roughly three orders of magnitude relative to AIMD and show near-linear scaling with system size. Despite these advances, clear performance gaps remain relative to DP and EAM, which are more than one order of magnitude faster than these FAMs. It is also noteworthy that these FAMs have substantial GPU memory requirements. For instance, MS-1M is limited to simulating systems of fewer than 4,000 atoms on an NVIDIA A100 GPU (40 GB). Even on the state-of-the-art NVIDIA H200 GPU (140 GB), the maximum tractable system size is approximately 10,000 atoms. This disadvantage may become a bottleneck for a few simulations relevant to Earth's core, such as solid–liquid coexistence calculations to compute liquidus temperatures and partition coefficients, which require ~10,000 atoms and nanosecond timescales [30,37,46,48]. It also limits rare-event simulations, such as inner-core nucleation [17,92-95], which are currently performed using EAM potentials. Consequently, distilling pretrained FAMs into efficient, task-specific "student" models and performing targeted fine-tuning for core-forming materials under extreme conditions emerge as promising strategies for Earth's core studies. Merging the broad chemical transferability of FAMs with the computational efficiency of specialized deep-learning potentials provides a critical pathway toward large-scale simulations required to probe complex structural and dynamical processes in Earth's core.

## V. Conclusion

In summary, by systematically benchmarking 17 FAMs under Earth's core conditions, we found that a few models can reproduce several structural and dynamical properties of Fe and Fe-rich alloys with core-relevant compositions. However, none of the tested models consistently reproduces all first-principles benchmarks. MatterSim, MACE, and DPA-3 provide reasonable descriptions of the EOS. The MatterSim models show more reliable descriptions of hcp-bcc-fcc stability and phonon behavior, whereas MACE substantially overestimates the stability of bcc Fe relative to hcp Fe. For iron alloys, the MatterSim models describe most binary liquid structures but fail for dilute oxygen-bearing liquids and superionic phases, while MACE performs better for oxygen-containing systems but shows artifacts in Fe-Si liquids. These limitations should stem from the lack of explicit treatment of thermal electronic excitations and the insufficient representation of dilute light-element configurations under high-P-T conditions in current training datasets. For the seven-element liquid, MS-5M gives the best agreement with AIMD, although several partial PCFs still show deviations. Overall, these results define both the current capabilities of FAMs for Earth's core materials and the future directions for distilling and fine-tuning models for accurate and efficient simulations of multicomponent systems under extreme core conditions.

**Acknowledgments**
Work at Xiamen University was supported by the National Natural Science Foundation of China (Grant No. 42550120). R.M.W acknowledges support from the Gordon & Betty Moore Foundation (doi.org/10.37807/GBMF12801). The Tan Kah Kee Supercomputing Center, as well as Shaorong

Fang and Tianfu Wu from the Information and Network Center of Xiamen University, are acknowledged for their support in GPU computing. Some simulations were performed on the Delta system at NCSA through allocation DMR180081 from the Advanced Cyberinfrastructure Coordination Ecosystem: Services & Support (ACCESS) program, which is supported by NSF Grants No. 2138259, No. 2138286, No. 2138307, No. 2137603, and No. 2138296.

**Data Availability**

The raw figure data and input files with technical settings for DFT and FAM calculations are openly available in Ref. [96]. These data can be used to reproduce the benchmarks and to evaluate future foundation atomistic models.

## Appendix

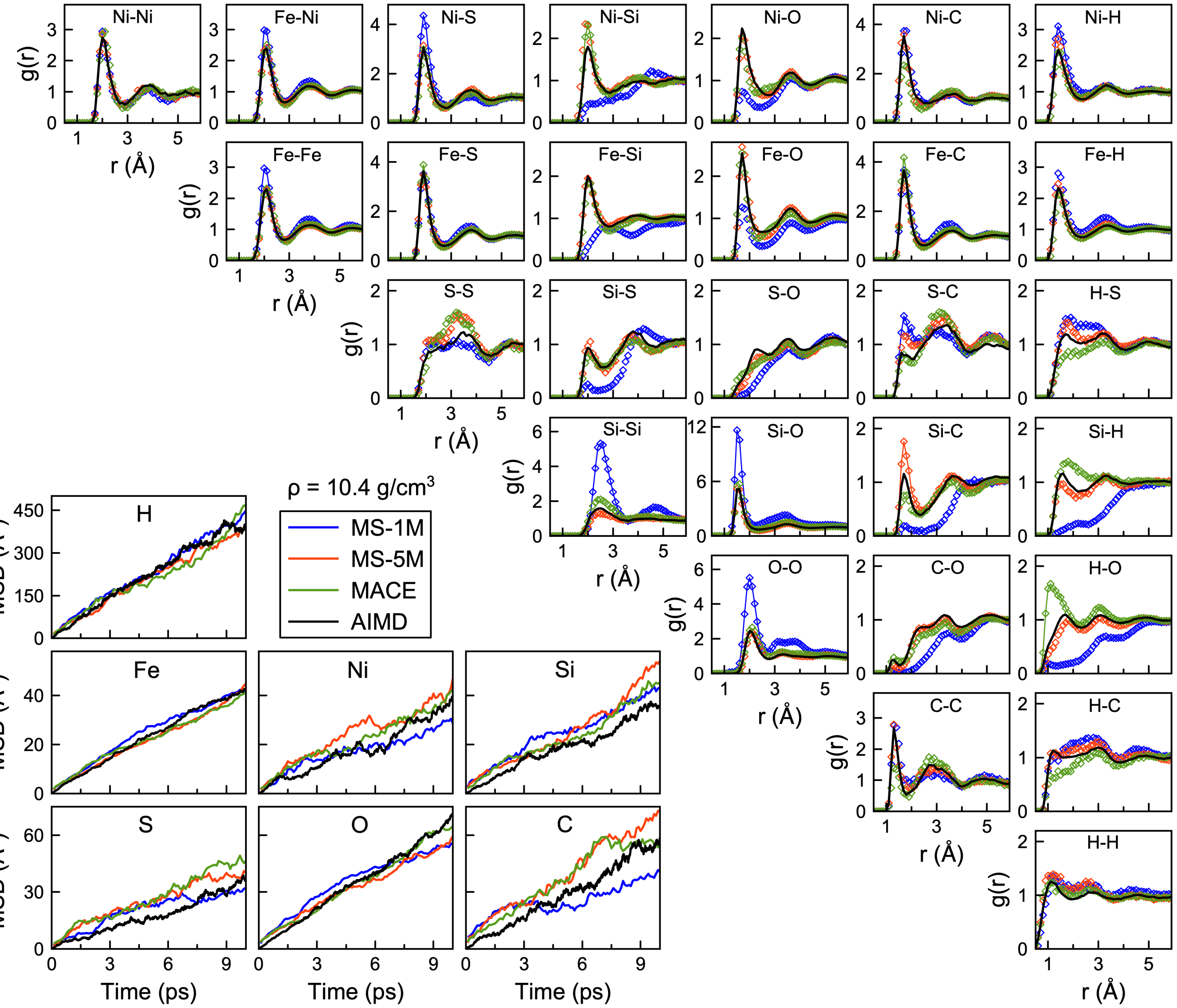

**FIG. A1** Partial PCF for all pairs and MSD for the seven elements for the $Fe_{150}Ni_{15}Si_{25}S_{15}H_{30}O_{80}C_{15}$ melts at 5800 K with a density of $10.4\ g/cm^3$, corresponding to 320 GPa in the AIMD simulation. Simulations with different models are performed at the same density, and the resulting pressures are listed in the legend.